\def\eck#1{\left\lbrack #1 \right\rbrack}
\def\eckk#1{\bigl[ #1 \bigr]}
\def\rund#1{\left( #1 \right)}
\def\abs#1{\left\vert #1 \right\vert}

\def\ave#1{\left\langle #1 \right\rangle}

\def\part#1#2{{\partial #1\over\partial #2}}

\def\A{{\cal A}}

\def\D{{\cal D}}
\def\I{{\cal I}}

\def\R{{\cal R}}

\def\d{{\rm d}}

\def\eps{{\epsilon}}

\def\vp{\varphi}
\def\vt{{\vartheta}}

\def\Real{{\rm I\mathchoice{\kern-0.70mm}{\kern-0.70mm}{\kern-0.65mm}%
  {\kern-0.50mm}R}}
\def\C{\rm C\kern-.42em\vrule width.03em height.58em depth-.02em
       \kern.4em}
\font \bolditalics = cmmib10
\def\bx#1{\leavevmode\thinspace\hbox{\vrule\vtop{\vbox{\hrule\kern1pt
        \hbox{\vphantom{\tt/}\thinspace{\bf#1}\thinspace}}
      \kern1pt\hrule}\vrule}\thinspace}

\def \vc #1{{\textfont1=\bolditalics \hbox{$\bf#1$}}}
{\catcode`\@=11
\gdef\SchlangeUnter#1#2{\lower2pt\vbox{\baselineskip 0pt \lineskip0pt
  \ialign{$\m@th#1\hfil##\hfil$\crcr#2\crcr\sim\crcr}}}
}

\def\ueber#1#2{{\setbox0=\hbox{$#1$}%
  \setbox1=\hbox to\wd0{\hss$\scriptscriptstyle #2$\hss}%
  \offinterlineskip
  \vbox{\box1\kern0.4mm\box0}}{}}

\def\bx#1{\leavevmode\thinspace\hbox{\vrule\vtop{\vbox{\hrule\kern1pt
        \hbox{\vphantom{\tt/}\thinspace{\bf#1}\thinspace}}
      \kern1pt\hrule}\vrule}\thinspace}

\def\SFB{{This work was supported by the ``Sonderforschungsbereich
375-95 f\"ur
Astro--Teil\-chen\-phy\-sik" der Deutschen For\-schungs\-ge\-mein\-schaft.}}
 
\magnification=\magstep1
\voffset= 0.0 true cm
\vsize=19.8 cm     
\hsize=13.5 cm
\hfuzz=2pt
\tolerance=500
\abovedisplayskip=3 mm plus6pt minus 4pt
\belowdisplayskip=3 mm plus6pt minus 4pt
\abovedisplayshortskip=0mm plus6pt
\belowdisplayshortskip=2 mm plus4pt minus 4pt
\predisplaypenalty=0
\footline={\tenrm\ifodd\pageno\hfil\folio\else\folio\hfil\fi}

\def\la{\mathrel{\hbox{\rlap{\hbox{\lower4pt\hbox{$\sim$}}}\hbox{$<$}}}}
\def\ga{\mathrel{\hbox{\rlap{\hbox{\lower4pt\hbox{$\sim$}}}\hbox{$>$}}}}

\def\utw{\smash{\rlap{\lower5pt\hbox{$\sim$}}}}
\def\udtw{\smash{\rlap{\lower6pt\hbox{$\approx$}}}}

\def\getsto{\mathrel{\hbox{\rlap{$\gets$}\hbox{\raise2pt\hbox{$\to$}}}}}
\def\lid{\mathrel{\hbox{\rlap{\hbox{\lower4pt\hbox{$=$}}}\hbox{$<$}}}}
\def\gid{\mathrel{\hbox{\rlap{\hbox{\lower4pt\hbox{$=$}}}\hbox{$>$}}}}
\def\sol{\mathrel{\hbox{\rlap{\hbox{\raise4pt\hbox{$\sim$}}}\hbox{$<$}}}
}
\def\sog{\mathrel{\hbox{\rlap{\hbox{\raise4pt\hbox{$\sim$}}}\hbox{$>$}}}
}
\def\lse{\mathrel{\hbox{\rlap{\hbox{\raise4pt\hbox{$<$}}}\hbox{$\simeq$}
}}}
\def\gse{\mathrel{\hbox{\rlap{\hbox{\raise4pt\hbox{$>$}}}\hbox{$\simeq$}
}}}
\def\grole{\mathrel{\hbox{\lower2pt\hbox{$<$}}\kern-8pt
\hbox{\raise2pt\hbox{$>$}}}}
\def\leogr{\mathrel{\hbox{\lower2pt\hbox{$>$}}\kern-8pt
\hbox{\raise2pt\hbox{$<$}}}}
\def\loa{\mathrel{\hbox{\rlap{\hbox{\lower4pt\hbox{$\approx$}}}\hbox{$<$
}}}}
\def\goa{\mathrel{\hbox{\rlap{\hbox{\lower4pt\hbox{$\approx$}}}\hbox{$>$
}}}}

%
%

\font\kleinhalbcurs=cmmib10 scaled 833
\font\eightrm=cmr8
\font\sixrm=cmr6
\font\eighti=cmmi8
\font\sixi=cmmi6
\skewchar\eighti='177 \skewchar\sixi='177
\font\eightsy=cmsy8
\font\sixsy=cmsy6
\skewchar\eightsy='60 \skewchar\sixsy='60
\font\eightbf=cmbx8
\font\sixbf=cmbx6
\font\eighttt=cmtt8
\hyphenchar\eighttt=-1
\font\eightsl=cmsl8
\font\eightit=cmti8

\font\bxf=cmbx10
  \mathchardef\Gamma="0100
  \mathchardef\Delta="0101
  \mathchardef\Theta="0102
  \mathchardef\Lambda="0103
  \mathchardef\Xi="0104
  \mathchardef\Pi="0105
  \mathchardef\Sigma="0106
  \mathchardef\Upsilon="0107
  \mathchardef\Phi="0108
  \mathchardef\Psi="0109
  \mathchardef\Omega="010A
\def\rahmen#1{\vskip#1truecm}
\def\begfig#1cm#2\endfig{\par
\setbox1=\vbox{\rahmen{#1}#2}%
\dimen0=\ht1\advance\dimen0by\dp1\advance\dimen0by5\baselineskip
\advance\dimen0by0.4true cm
\ifdim\dimen0>\vsize\pageinsert\box1\vfill\endinsert
\else
\dimen0=\pagetotal\ifdim\dimen0<\pagegoal
\advance\dimen0by\ht1\advance\dimen0by\dp1\advance\dimen0by1.4true cm
\ifdim\dimen0>\vsize
\topinsert\box1\endinsert
\else\vskip1true cm\box1\vskip4true mm\fi
\else\vskip1true cm\box1\vskip4true mm\fi\fi}
\def\figure#1#2{\smallskip\setbox0=\vbox{\noindent\petit{\bf Fig.\ts#1.\
}\ignorespaces #2\smallskip
\count255=0\global\advance\count255by\prevgraf}%
\ifnum\count255>1\box0\else
\centerline{\petit{\bf Fig.\ts#1.\ }\ignorespaces#2}\smallskip\fi}


\def\begtab#1cm#2\endtab{\par
\ifvoid\topins\midinsert\vbox{#2\rahmen{#1}}\endinsert
\else\topinsert\vbox{#2\kern#1true cm}\endinsert\fi}
\def\rahmen#1{\vskip#1truecm}
\def\begpet{\vskip6pt\bgroup\petit}
\def\endpet{\vskip6pt\egroup}
\def\begref{\par\bgroup\petit
\let\it=\rm\let\bf=\rm\let\sl=\rm\let\INS=N}
\def\petit{\def\rm{\fam0\eightrm}%
\textfont0=\eightrm \scriptfont0=\sixrm \scriptscriptfont0=\fiverm
 \textfont1=\eighti \scriptfont1=\sixi \scriptscriptfont1=\fivei
 \textfont2=\eightsy \scriptfont2=\sixsy \scriptscriptfont2=\fivesy
 \def\it{\fam\itfam\eightit}%
 \textfont\itfam=\eightit
 \def\sl{\fam\slfam\eightsl}%
 \textfont\slfam=\eightsl
 \def\bf{\fam\bffam\eightbf}%
 \textfont\bffam=\eightbf \scriptfont\bffam=\sixbf
 \scriptscriptfont\bffam=\fivebf
 \def\tt{\fam\ttfam\eighttt}%
 \textfont\ttfam=\eighttt
 \normalbaselineskip=9pt
 \setbox\strutbox=\hbox{\vrule height7pt depth2pt width0pt}%
 \normalbaselines\rm
\def\vec##1{\setbox0=\hbox{$##1$}\hbox{\hbox
to0pt{\copy0\hss}\kern0.45pt\box0}}}%
\let\ts=\thinspace
%
\font \tafontt=     cmbx10 scaled\magstep2
\font \tafonts=     cmbx7  scaled\magstep2
\font \tafontss=     cmbx5  scaled\magstep2
\font \tamt= cmmib10 scaled\magstep2
\font \tams= cmmib10 scaled\magstep1
\font \tamss= cmmib10
\font \tast= cmsy10 scaled\magstep2
\font \tass= cmsy7  scaled\magstep2
\font \tasss= cmsy5  scaled\magstep2
\font \tasyt= cmex10 scaled\magstep2
\font \tasys= cmex10 scaled\magstep1
\font \tbfontt=     cmbx10 scaled\magstep1
\font \tbfonts=     cmbx7  scaled\magstep1
\font \tbfontss=     cmbx5  scaled\magstep1
\font \tbst= cmsy10 scaled\magstep1
\font \tbss= cmsy7  scaled\magstep1
\font \tbsss= cmsy5  scaled\magstep1

\newbox\chsta\newbox\chstb\newbox\chstc
\def\centerpar#1{{\advance\hsize by-2\parindent
\rightskip=0pt plus 4em
\leftskip=0pt plus 4em
\parindent=0pt\setbox\chsta=\vbox{#1}%
\global\setbox\chstb=\vbox{\unvbox\chsta
\setbox\chstc=\lastbox
\line{\hfill\unhbox\chstc\unskip\unskip\unpenalty\hfill}}}%
\leftline{\kern\parindent\box\chstb}}
 \def \chap#1{
    \vskip24pt plus 6pt minus 4pt
    \bgroup
 \textfont0=\tafontt \scriptfont0=\tafonts \scriptscriptfont0=\tafontss
 \textfont1=\tamt \scriptfont1=\tams \scriptscriptfont1=\tamss
 \textfont2=\tast \scriptfont2=\tass \scriptscriptfont2=\tasss
 \textfont3=\tasyt \scriptfont3=\tasys \scriptscriptfont3=\tenex
     \baselineskip=18pt
     \lineskip=18pt
     \raggedright
     \pretolerance=10000
     \noindent
     \tafontt
     \ignorespaces#1\vskip7true mm plus6pt minus 4pt
     \egroup\noindent\ignorespaces}%
 \def \sec#1{
     \vskip25true pt plus4pt minus4pt
     \bgroup
 \textfont0=\tbfontt \scriptfont0=\tbfonts \scriptscriptfont0=\tbfontss
 \textfont1=\tams \scriptfont1=\tamss \scriptscriptfont1=\kleinhalbcurs
 \textfont2=\tbst \scriptfont2=\tbss \scriptscriptfont2=\tbsss
 \textfont3=\tasys \scriptfont3=\tenex \scriptscriptfont3=\tenex
     \baselineskip=16pt
     \lineskip=16pt
     \raggedright
     \pretolerance=10000
     \noindent
     \tbfontt
     \ignorespaces #1
     \vskip12true pt plus4pt minus4pt\egroup\noindent\ignorespaces}%
 \def \subs#1{
     \vskip15true pt plus 4pt minus4pt
     \bgroup
     \bxf
     \noindent
     \raggedright
     \pretolerance=10000
     \ignorespaces #1
     \vskip6true pt plus4pt minus4pt\egroup
     \noindent\ignorespaces}%
 \def \subsubs#1{
     \vskip15true pt plus 4pt minus 4pt
     \bgroup
     \bf
     \noindent
     \ignorespaces #1\unskip.\ \egroup
     \ignorespaces}
\def\footnoterule{\kern-3pt\hrule width 2true cm\kern2.6pt}
\newcount\footcount \footcount=0
\def\advftncnt{\advance\footcount by1\global\footcount=\footcount}
\def\fonote#1{\advftncnt$^{\the\footcount}$\begingroup\petit
       \def\textindent##1{\hang\noindent\hbox
       to\parindent{##1\hss}\ignorespaces}%
\vfootnote{$^{\the\footcount}$}{#1}\endgroup}

\newcount\sterne
\outer\def\byebye{\bigskip\typeset
\sterne=1\ifx\speciali\undefined\else
\bigskip Special caracters created by the author
\loop\smallskip\noindent special character No\number\sterne:
\csname special\romannumeral\sterne\endcsname
\advance\sterne by 1\global\sterne=\sterne
\ifnum\sterne<11\repeat\fi
\vfill\supereject\end}
\def\typeset{\centerline{\petit This article was processed by the author
using the \TeX\ Macropackage from Springer-Verlag.}}
 
\voffset=0pt

\def\T{{\cal T}}
\chap{The cosmological lens equation and the equivalent single-plane
gravitational lens}
\medskip
\centerline{\bf Peter Schneider}
\centerline{\bf Max-Planck-Institut f\"ur Astrophysik}
\centerline{\bf Postfach 1523}
\centerline{\bf D-85740 Garching, Germany}
\bigskip
\sec{Abstract}
The gravitational lens equation resulting from a single (non-linear)
mass concentration (the main lens) plus inhomogeneities of the
large-scale structure is shown to be strictly equivalent to the
single-plane gravitational lens equation without the cosmological
perturbations. The deflection potential (and, by applying the Poisson
equation, also the mass distribution) of the equivalent single-plane
lens is derived. If the main lens is described by elliptical
isopotential curves plus a shear term, the equivalent single-plane
lens will be of the same form.  Due to the equivalence shown, the
determination of the Hubble constant from time delay measurements is
affected by the same mass-sheet invariance transformation as for the
single-plane lens. If the lens strength is fixed (e.g., by measuring
the velocity dispersion of stars in the main lens), the determination
of $H_0$ is affected by inhomogeneous matter between us and the lens.
The orientation of the mass
distribution relative
to the image positions is the same for the cosmological lens
situation and the single-plane case. In particular this implies that
cosmic shear cannot account for a misalignment of the observed galaxy
orientation relative to the best-fitting lens model.

\sec{1 Introduction}
Light from distant sources propagates through the matter
inhomogeneities of our Universe; light bundles are deflected and
distorted during their propagation (e.g., Gunn 1967a,b). This
gravitational light deflection can be used to infer statistical
properties of the large-scale matter distribution in the Universe, as
was pointed out by Miralda-Escud\'e (1991),
Blandford et al. (1991), Kaiser (1992, 1996), Villumsen
(1996), Bernardeau, van Waerbeke \& Mellier (1996), and Jain \& Seljak
(1996), by measuring the mean ellipticity of the images of distant
galaxies; assuming that galaxies are intrinsically randomly oriented,
any net ellipticity is then attributed to the propagation. Estimates of
the expected effect have both been done numerically (e.g.,
Jaroszy\'nski, Park \& Paczy\'nski 1990,
Jaroszy\'nski 1991, 1992, Bartelmann \& Schneider 1991, Wambsganss et
al.\ts 1996) and
analytically, using the linear or fully non-linear evolution of the
power spectrum of density fluctuations (Seljak 1996, Jain \& Seljak
1996, and references therein). 

This cosmic shear acts of course also on light bundles corresponding
to gravitational lens systems such as multiply-imaged QSOs. Since the
cosmic shear is a stochastic process, its value in the direction of a
gravitational lens system is unknown and has to be determined from
lens modeling. By that -- as we shall see below -- the cosmic shear
cannot be distinguished from 
locally generated shear, i.e., by a group of galaxies or a cluster
close to the main lens. 
In addition to the shear, the large-scale matter
inhomogeneities can also produce a convergence (which can have
either sign). This convergence which is indistinguishable from a
convergence caused by the local environment of the main lens, affects
the determination of the Hubble constant (Kayser \& Refsdal 1983,
Gorenstein, Falco \& Shapiro 1988). The primary effect here is that
matter inhomogeneities between us and the lens change the
angular-diameter distance to the lens, which essentially is the
quantity determined from a measurement of the time delay (Narayan
1991). 

Surpi, Harari \& Frieman (1996) and 
Bar-Kana (1996, hereafter B-K96) have investigated the effect of
large-scale structure on multiply imaged sources. In B-K96, the effect
of cosmic shear added to a given gravitational lens was considered and
shown to affect the observable image positions and flux
ratios. However, since the mass model for the gravitational lens is
obtained by fitting observable image properties with model
predictions, the addition of a cosmic shear implies that the adopted
lens model is modified such as to reproduce the observables best.
Therefore if one wants to investigate the effect of cosmic shear on
multiple image lens systems, the combined effect of adding cosmic
shear and modifying the lens model accordingly must be studied.

The present paper
considers some aspects of B-K96 in more detail. The main results to be
shown are as follows: (1) B-K96 derives a lens equation including
cosmic shear which is {\it formally} identical to the (multiple-plane)
generalized quadrupole gravitational lens equation (Kovner 1987;
Schneider, Ehlers \& Falco 1992; hereafter SEF, Chap.\ts 9). B-K96
shows that the analogue of the `telescope matrix' (Kovner 1987) is
symmetric if the effects of the large-scale matter inhomogeneities are
considered to first order. It will be shown below that this matrix is
manifestly symmetric to all orders. This is not unimportant: whereas
the rms value of the cosmic shear is below 10\%, it obeys non-Gaussian
statistics, and considerably higher values are probably not rare.  (2)
The symmetry of this matrix is then used to show that the cosmic lens
equation is fully equivalent to a single-plane gravitational lens
equation, and the equivalent single-plane matter distribution (or,
equivalently, the single-plane deflection potential) is obtained. (3)
If a family of lens models is considered where the main lens consists
of a linear term (a shear matrix) plus a main lens with elliptical
potential, then the equivalent single-plane lens is contained in this
family of lens models. Note that this kind of lens models has been
applied to observed lens systems, and it appears that an (external)
shear in addition to the ellipticity of the lens galaxy is needed in
most cases where enough observational constraints are available (e.g.,
four-image systems) -- see Keeton, Kochanek \& Seljak (1996), Witt \&
Mao (1997).  (4) The observed image positions relative to the
observable major axis of the matter distribution in the main lens is
the same in the cosmic lens situation and in the equivalent
single-plane situation, in contrast to the possibility indicated by
Blandford \& Kundi\'c (1996) who suspected possible misalignment of the
images with respect to the observed orientation of the galaxy. 
\vfill\eject

\sec{2 Light propagation in an inhomogeneous Universe} 
In this section we closely follow the paper by Seitz, Schneider \&
Ehlers (1994; hereafter SSE); the reader is referred to this paper for
details. 

\subs{2.1 General propagation equations}
Consider a fiducial light ray ending at the observer at
redshift $z=0$. The corresponding null geodesic is denoted by
$\gamma^\mu_0(\lambda)$, where $\lambda$ is an affine parameter along
the ray. We choose $\lambda$ such that $\lambda=0$ at the observer,
and that locally $\lambda$ coincides with the proper
distance. Consider a neighbouring light ray
$\gamma^\mu(\lambda;\vc\theta)$ which at the observer propagates at an
angle $\vc\theta$ relative to the central ray. The transverse
component of the separation vector
$\xi^\mu(\lambda;\vc\theta)
=\gamma^\mu(\lambda;\vc\theta)-\gamma^\mu_0(\lambda)$ is (essentially)
a two-dimensional vector which shall be denoted by
$\vc\xi(\lambda;\vc\theta)$ (for an exact definition, see
SSE). Provided $\vc\theta$ is sufficiently small,
$\vc\xi(\lambda;\vc\theta)$ will depend linearly on $\vc\theta$; it
satisfies the Jacobi differential equation
$$
\vc\xi''(\lambda;\vc\theta)=
\T(\lambda)\vc\xi(\lambda;\vc\theta)\quad,
\eqno (2.1)
$$
where $\T(\lambda)$ is the optical tidal matrix evaluated at the
position $\gamma^\mu_0(\lambda)$ of the fiducial ray at the affine
parameter $\lambda$; see SSE for the general definition of the optical
tidal matrix. A prime denotes differentiation with respect to
$\lambda$. Equivalently, we can write
$$
\vc\xi(\lambda;\vc\theta)=\D(\lambda)\vc\theta\quad,
\eqno (2.2)
$$
which after insertion into (2.1) yields an equation for the matrix $\D$,
$$
\D''(\lambda)=\T(\lambda)\,\D(\lambda)\quad .
\eqno (2.3)
$$
From the requirement that $\lambda$ agrees locally with the proper
distance, the initial conditions for $\vc\xi$ are $\vc\xi(0)=\vc 0$,
$\vc\xi'(0)=\vc\theta$, or equivalently, $\D(0)=0$, $\D'(0)=\I$, where
$\I$ denotes the unit matrix.

Suppose the matter distribution in the Universe is characterized by a
homogeneous matter density $\bar\rho(z)=(1+z)^3\Omega_0 \rho_{\rm
cr}$, with critical density $\rho_{\rm cr}=3H_0^2/(8\pi G)$; in that
case, the optical tidal matrix $\T$ is proportional to the unit matrix
and is given by
$$
\T(\lambda)=-{3\over 2}\rund{H_0\over c}^2 
\Omega_0\,(1+z)^5\,\I\quad .
\eqno (2.4)
$$
The relation between the affine parameter and the redshift $z$ reads
$$
\d\lambda={c\over H_0}{\d z\over (1+z)^3
\sqrt{1+z\Omega_0-\Omega_\Lambda\eck{1-(1+z)^{-2}}}}\quad ,
\eqno (2.5)
$$
where $\Omega_\Lambda= \Lambda/(3 H_0^2)$ is the density parameter
associated with the cosmological constant $\Lambda$. In this case of a
homogeneous Universe, the matrix $\D(\lambda)$ remains proportional to
the unit matrix, $\D(\lambda)=D(\lambda)\,\I$, which implies an
isotropic propagation of light bundles; i.e., the cross-section of
initially circular light bundles remains circular. The
angular-diameter-distance $D$ as a function of $\lambda$ or $z$ can
then be obtained from the 
differential equation (2.3) for the $\D$, using (2.4). 

Adding density perturbations $\delta\rho$ to the Universe, the optical
tidal matrix changes to
$$
\T_{ij}(\lambda)=-{3\over2}\rund{H_0\over c}^2\Omega_0
(1+z)^5\,\delta_{ij} -{(1+z)^2\over c^2}
\rund{2\Phi_{,ij}+\delta_{ij}\Phi_{,33}}\quad ,
\eqno (2.6)
$$
where the local coordinates are such that the light ray propagates in
the $x_3$-direction. The gravitational potential $\Phi$ is related to
the density perturbations via the Poisson equation $\nabla^2\Phi=4\pi
G \delta\rho$ (where the differential operator is taken with respect
to local proper coordinates) which can be justified provided the
density perturbations have typical scales much smaller than the radius
of curvature of the Universe, i.e., if that scale is much less than
the Hubble radius $c/H_0$. In general, the matrix $\D(\lambda)$ now
attains shear components, and initially circular light bundles obtain
elliptical cross sections.

Finally, we add to the density perturbations a non-linear
gravitational lens at redshift $z_{\rm d}$, described by its surface
mass density $\Sigma(\vc\xi)$, where $\vc\xi$ is the proper distance
vector in the lens plane, i.e., a plane perpendicular to the light
rays under consideration. If the fiducial light ray is chosen such
that it traverses the lens plane at $\vc\xi=\vc 0$, then the
separation vector $\vc\xi(\lambda)$ becomes
$$
\vc\xi(\lambda)=\cases{\D(\lambda) \vc\theta &for $\lambda\le
\lambda_{\rm d}$ \cr
\D(\lambda) \vc\theta -\hat\D(\lambda)\eck{
\vc\alpha(\vc\xi(\lambda_{\rm d})) -\vc\alpha(\vc 0)}
&for $\lambda > \lambda_{\rm d}$ \cr}\quad ,
\eqno (2.7)
$$
where $\vc\alpha(\vc\xi)$ is the deflection angle a light ray
undergoes at position $\vc\xi$,
$$
\vc\alpha(\vc\xi)={4G\over c^2}\int_{\Real^2}\d^2\xi'\;
\Sigma(\vc\xi')\,{\vc\xi-\vc\xi' \over \abs{\vc\xi-\vc\xi'}^2}\quad ;
\eqno (2.8)
$$
the fact that $\vc\xi(\lambda)$ satisfies the propagation equation
(2.1) for all $\lambda\ne\lambda_{\rm d}$ with the optical tidal
matrix (2.6) implies that $\hat\D(\lambda)$ also satisfies the
differential equation (2.3). Eq.(2.7) is derived in Sect.\ts 4.4 of
SSE, where it is also shown that the requirement that the local change
of direction of the light ray in the lens plane agrees with the
deflection angle implies that the initial conditions for $\hat\D$ are
$\hat\D(\lambda_{\rm d})=0$, $\hat\D'(\lambda_{\rm d})=(1+z_{\rm
d})\I$.

\subs{2.2 The cosmological lens equation}
In the case of an isolated mass concentration acting as a
gravitational lens, there always exists at least one point where the
deflection angle vanishes; this can be easily shown with the
Poincar\'e's index theorem (see, e.g., Sect.\ts 5.4.1 of SEF). We
shall assume that the origin in the lens plane -- and thus the
fiducial ray $\gamma^\mu_0$ -- is chosen such that
$\vc\alpha(\vc 0)=\vc 0$.

Consider sources at redshift $z_{\rm s}$, corresponding to the affine
parameter $\lambda_{\rm s}$. Let $\vc\eta\equiv \vc\xi(\lambda_{\rm
s})$, and, as before, $\vc\xi\equiv \vc\xi(\lambda_{\rm
d})$. The lens equation in an inhomogeneous Universe then becomes
$$
\vc\eta=\D(\lambda_{\rm s})\D^{-1}(\lambda_{\rm d})\vc\xi-
\hat\D(\lambda_{\rm s})\vc\alpha(\vc\xi)\quad ,
\eqno (2.9)
$$
where we have assumed that the lensing effect by the perturbations
$\delta\rho$ is sufficiently small so that no caustic points are
caused by these perturbations themselves. This implies that the
matrices $\D(\lambda)$ and $\hat\D(\lambda)$ are nowhere singular and
thus can be inverted. In (2.9), we have written
$\vc\xi=\D(\lambda_{\rm d})\vc\theta$ as independent variable; the
non-singularity requirement for $\D$ implies that there is a
one-to-one relation between $\vc\xi$ and $\vc\theta$. Eq.\ts(2.9) maps
a vector $\vc\xi$ from the lens plane into a vector $\vc\eta$ in the
source plane, just as in standard gravitational lens theory; the
difference between (2.9) and the standard lens equation is that in the
latter case, the matrices $\D$ and $\hat\D$ are proportional to the
unit matrix, i.e., they effectively become scalars. These scalars are
the angular diameter-distances, which are uniquely defined in terms of
redshifts in a homogeneous Universe. In an inhomogeneous Universe, the
matrices $\D$ and $\hat\D$ attain shear components, so that the
angular diameter-distances are no longer isotropic; moreover, since
the density perturbations $\delta\rho$ form a random field, the
optical tidal matrix 
along the direction to a gravitational
lens has random components, and so there is basically no hope to obtain
enough knowledge about this matter distribution to determine
$\D(\lambda_{\rm d})$, $\D(\lambda_{\rm s})$, and $\hat\D(\lambda_{\rm
s})$ for a given lens system. For notational convenience, we define 
$$
\D_{\rm d}\equiv \D(\lambda_{\rm d}) \quad ;\quad 
\D_{\rm s}\equiv \D(\lambda_{\rm s}) \quad ;\quad
\hat\D_{\rm s}\equiv \hat\D(\lambda_{\rm s})\quad .
\eqno (2.10)
$$

Furthermore it should be noted that these matrices are in general not
symmetric, so that the cosmic perturbations induce a rotational
component to the matrices $\D$ and $\hat\D$ (this has been called
`twist' in SSE). However, this twist is unobservable: let $\vc\beta =
\hat\D^{-1}(\lambda_{\rm s}) \vc\eta$; then,
$$ 
\vc\beta=\hat\D^{-1}_{\rm s}
\D_{\rm s}\D^{-1}_{\rm d}\vc\xi
-\vc\alpha(\vc\xi)=: C_\xi \vc\xi -\vc\alpha(\vc\xi)\quad .
\eqno (2.11)
$$
As we shall show in the Appendix, the matrix $C_\xi$ multiplying
$\vc\xi$ in (2.11) is symmetric, so that the matrix
$\partial\vc\beta/\partial\vc\xi$ is symmetric. The symmetry of
$C_\xi$ is in complete analogy to the symmetry of the `telescope
matrix' introduced by Kovner (1987) in the frame of a multiple
deflection gravitational lens system with at most one non-linear
deflector; see also SEF, Sect.\ts 9.3. The explicit algebraic proof
for the symmetry of the `telescope matrix' has been given in Seitz \&
Schneider (1994). We also note in passing that the magnification
theorem (Schneider 1984; Seitz \& Schneider 1992) is still satisfied:
at least one of the images formed by the gravitational lens of any
source is brighter than the source would appear in the same direction
(i.e., with the same optical tidal matrix) in the absence of the lens,
due to the non-negativity of $\Sigma(\vc\xi)$. However, it should be
noted that the source can appear fainter than the same source at the
same redshift would appear in the corresponding homogeneous Universe,
since $\delta\rho$ can be negative; indeed, $\ave{\delta\rho}=0$ by
definition.

We write the lens equation (2.11) in terms of the angle $\vc\theta$,
by multiplying (2.11) from the left by $\D^{\rm T}_{\rm d}$,
and by defining $\vc\eta'=\D^{\rm T}_{\rm d}\vc\beta$:
$$
\vc\eta'=\D^{\rm T}_{\rm d}\hat\D^{-1}_{\rm s}
\D_{\rm s}\vc\theta - \D^{\rm T}_{\rm d}
\vc\alpha\rund{\D_{\rm d}\vc\theta}
=: C_\theta \vc\theta -\D^{\rm T}_{\rm d}
\vc\alpha\rund{\D_{\rm d}\vc\theta} \quad .
\eqno (2.12)
$$
Since $C_\theta=\D^{\rm T}_{\rm d} C_\xi \D_{\rm d}$, the
symmetry of $C_\xi$ implies that $C_\theta$ is also symmetric.

\sec{3 Relation to observables}
The `source position' $\vc\eta'$ is obtained from the `true' source
position $\vc\eta$ through a linear transformation. However, this
transformation does not affect the observables. Let $\vc\theta^i$ be
the observed image positions, $1\le i\le N$, then the lens equation
predicts that they have to satisfy
$$
C_\theta \rund{\vc\theta^i-\vc\theta^j}=\D_{\rm d}^{\rm
T}\eck{\vc\alpha\rund{\D_{\rm d}\vc\theta^i}
-\vc\alpha\rund{\D_{\rm d}\vc\theta^j}}\quad ,
\eqno (3.1)
$$
for $i\ne j$, a relation not containing $\vc\eta'$. Secondly, whereas the
magnification of the images depends on the linear transformation in
the source plane, the observable magnification ratios are unaffected
by it. What can be observed are the relative magnification matrices
$\A_{ij}$ between images,
$$
\A_{ij}=\rund{\partial \vc\eta'(\vc\theta^i)\over\partial\vc\theta}
\rund{\partial \vc\eta'(\vc\theta^j)\over\partial\vc\theta}^{-1}\quad,
\eqno (3.2)
$$
which are unchanged by a linear transformation in the source
plane. Note that
$$
{\partial \vc\eta'\over\partial \vc\theta}=C_\theta-\D^{\rm T}_{\rm d}
U\rund{\D_{\rm d}\vc\theta}\D_{\rm d}\quad ,
\eqno (3.3)
$$
where
$$
U(\vc\xi)={\partial \vc\alpha(\vc\xi)\over\partial\vc\xi}\quad .
\eqno (3.4)
$$
Since $U$ is symmetric, so is ${\partial \vc\eta'/\partial
\vc\theta}$. 

We next consider the time delay function $T$. For that we first note
that the potential part of the time delay is (Cooke \& Kantowski 1975)
$$
c T_{\rm pot}=(1+z_{\rm d}) \Psi(\vc\xi) \quad ,
\eqno (3.5)
$$
where
$$
\Psi(\vc\xi)={4G\over c^2}\int_{\Real^2}\d^2\xi'\;
\Sigma(\vc\xi')\,\ln\rund{ \abs{\vc\xi-\vc\xi'}\over\xi_*}\quad ,
\eqno (3.6)
$$
is the deflection potential, and $\xi_*$ is an arbitrary length
scale; changing $\xi_*$ changes $\Psi$ by an additive constant which
does therefore not affect the measureable time delays. From Fermat's
principle (Schneider 1985, Kovner 1990) one knows that the lens
equation is equivalent 
to $\nabla T=\vc 0$, which determines $T$ up to an affine
transformation. The multiplicative constant is determined by that of
$T_{\rm pot}$, so that
$$
c T=(1+z_{\rm d})\eck{\vc\theta\cdot(C_\theta \vc\theta)/2
-\vc\eta'\cdot\vc\theta -\Psi(\D_{\rm d}\vc\theta)}\quad ,
\eqno (3.7)
$$
and noting that $\partial\Psi/\partial\vc\theta
=\D_{\rm d}^{\rm T}\vc\alpha(\D_{\rm d}\vc\theta)$, one sees that 
$\partial T/\partial\vc\theta =\vc 0$ is equivalent to (2.12). The
time delay between any pair of images is then $\Delta
t=T\rund{\vc\theta^i} - T\rund{\vc\theta^j}$. Whereas $T$ as written
in the form (3.7) contains the source position, one notes that
$\vc\eta'$ is substituted in terms of the observed image positions
using the lens equation (2.12). Therefore, the linear transformation
in the source plane does not affect the calculation of the time delay.

\sec{4 The equivalent single-plane lens}
We shall now show that the lens equation (2.12) is equivalent to a
single-plane lens situation without `cosmic shear'. The single-plane
lens equation reads
$$
\tilde\vc\eta=D_{\rm s}\vc\theta-D_{\rm ds}\tilde\vc\alpha(D_{\rm
d}\vc\theta)\quad,
\eqno (4.1)
$$
where $D_{\rm d}$, $D_{\rm ds}$,
and $D_{\rm s}$ are the angular diameter distances
in a homogeneous Universe, as described in Sect.\ts 2. We characterize
quantities in the single-plane lens with a tilde, to distinguish them
from those of the cosmological lens equation. As before,
$\tilde\vc\alpha(\tilde\vc\xi)$ can be obtained from a deflection
potential
$$
\tilde\vc\alpha(\tilde\vc\xi)={\d \tilde\Psi(\tilde\vc\xi)\over \d
\tilde\vc\xi} \quad .
\eqno (4.2)
$$

\noindent
{\bf Theorem:} For every mass distribution $\Sigma(\vc\xi)$ and
matrices $\D_{\rm d}$, $\D_{\rm s}$ and $\hat\D_{\rm s}$ giving rise
to the cosmological lens equation (2.12), there exists a mass
distribution $\tilde\Sigma(\tilde\vc\xi)$ for which the single-plane
lens equation (4.1) yields the same observables.

This will be shown by construction. To simplify notation we define the
matrices
$$
\R_{\rm d}={\D_{\rm d}\over D_{\rm d}} \quad ;\quad
\R_{\rm s}={\D_{\rm s}\over D_{\rm s}} \quad ;\quad
\R_{\rm ds}={\hat\D_{\rm s}\over D_{\rm ds}}\quad,
\eqno (4.3)
$$
which describe the deviation of the light propagation in an
inhomogeneous Universe from that in the homogeneous one; i.e., in a
homogeneous Universe these three matrices would become the unit
matrix. Define the deflection potential
$$
\tilde\Psi(\tilde\vc\xi):=\Psi\rund{\R_{\rm d}\tilde\vc\xi}
+{1\over 2}\tilde\vc\xi\cdot\rund{B\tilde\vc\xi}\quad ,
\eqno (4.4)
$$
where $B$ is the matrix
$$
B={D_{\rm s}\over D_{\rm d}}\rund{\R_{\rm d}^{\rm T} \R_{\rm ds}^{-1}
\R_{\rm s} -\I}\quad ;
\eqno (4.5)
$$
then the deflection angle becomes, according to (4.2),
$$
\tilde\vc\alpha(\tilde\vc\xi)
=\R_{\rm d}^{\rm T}\vc\alpha(\R_{\rm d}\tilde\vc\xi)
+B\tilde\vc\xi\quad .
\eqno (4.6)
$$
This is now inserted into the lens equation (4.1) to yield
$$
\tilde\vc\eta=D_{\rm s}\vc\theta -D_{\rm ds}\R_{\rm d}^{\rm T}
\vc\alpha(\D_{\rm d}\vc\theta)+D_{\rm d} B\vc\theta\quad .
\eqno (4.7)
$$
If we now define $\tilde\vc\eta'\equiv (D_{\rm d}/D_{\rm
ds})\tilde\vc\eta$, then the lens equation (4.7) becomes after
multiplication with $(D_{\rm d}/D_{\rm ds})$
$$
\tilde\vc\eta'=C_\theta \vc\theta-\D_{\rm d}^{\rm T}\vc\alpha(\D_{\rm
d}\vc\theta)\quad ,
\eqno (4.8)
$$
which is obviously the same equation as (2.12), except for a linear
transformation of the source coordinates. In particular, this
single-plane lens yields the same relations (3.1), (3.2) and (3.7)
between the image positions, the relative magnification matrices and
the time delay function as the cosmological lens mapping. The surface
mass density corresponding to the equivalent single-plane lens is
obtained from (4.4) via Poisson's equation. The first term in (4.4)
then yields a contribution to $\tilde\Sigma(\tilde\vc\xi)$ which is
obtained from $\Sigma(\vc\xi)$ in a non-trivial way, although the
corresponding potential contributions are related by a simple linear
transformation. The second term in (4.4) yields a uniform surface mass
density. Note that both contributions are not guaranteed to be
non-negative.

If we use the angular source position
$\tilde\vc\beta=\tilde\vc\eta/D_{\rm s}$, the lens equation takes the
more familiar form
$$
\tilde\vc\beta=\R_{\rm d}^{\rm T} \R_{\rm ds}^{-1} \R_{\rm s}\vc\theta
-{D_{\rm ds}\over D_{\rm s}}\R_{\rm d}^{\rm T}
\vc\alpha(\D_{\rm d}\vc\theta)
=:\tilde\Gamma \vc\theta-{D_{\rm ds}\over D_{\rm s}}\R_{\rm d}^{\rm T}
\vc\alpha(\D_{\rm d}\vc\theta)\quad ,
\eqno (4.9)
$$
so that in fact the equivalent single-plane lens consists of the shear
matrix $\tilde\Gamma$ plus a term corresponding to the `main lens'. 
Since $\tilde\Gamma$ is proportional to $C_\theta$ it is also symmetric.

In general, the trace of $\tilde\Gamma$ is different from 2, which means
that the shear matrix contains an equivalent surface mass density term
(which is not necessarily positive). As was pointed out by Gorenstein,
Falco \& Shapiro (1988), adding a homogeneous surface mass density to
a lens and at the same time rescaling the lens strength leaves all
observables invariant except the time delay. This mass-sheet
degeneracy is therefore also present in the single-plane lens and is
not a novel feature of the cosmological lens equation.

\sec{5 Special case: Elliptical isopotentials}
We now consider the case that the lens is described by elliptical
isopotential curves, so that
$$
\Psi(\vc\xi)=F(u)\quad,
\eqno (5.1)
$$
with
$$
u=\xi^2\rund{1-\eps \cos\eckk{2\rund{\vp-\vt}}}\quad ;
\eqno (5.2)
$$
here, $\vc\xi$ has been expressed in polar coordinates
$\vc\xi=\xi(\cos\vp,\sin\vp)$, $\eps$ describes the ellipticity of the
potential, and $\vt$ is the direction of the major axis of the
potential. This kind of lens model has been introduced by Blandford \&
Kochanek (1987) and investigated in detail by several authors (e.g.,
Kassiola \& Kovner 1993). The corresponding mass distribution has no
elliptical isodensity curves, and depending on the shape of the radial
profile $F(u)$ and the ellipticity $\eps$, dumb bell shaped isodensity
curves or even negative surface mass densities are obtained from
$\Psi$. Leaving aside these potential difficulties, models of the form
(5.1) have often been used for modeling gravitational lens systems. 
Whereas the surface mass density is not elliptical, one can still
define a major and minor axis, which coincide with the major and minor
axes of $\Psi$. 

Expressed in terms of the observable $\vc\theta$, $\Psi(\D_{\rm
d}\vc\theta)$, the observed major axis of the lens does not coincide
with $\vt$, but is rotated by the matrix $\D_{\rm d}$. But the same is
true for the equivalent single-plane lens, see (4.4). Note that the
first term in (4.4) is again an elliptical potential in
$\tilde\vc\xi$, however with different ellipticity and orientation, and
a modified radial profile. However,
the matrix which
rotates $\tilde\Psi(D_{\rm d}\vc\theta)$ with respect to $\Psi$ is the
same as that rotating the intrinsic lens direction $\vt$ into the
observed one. Hence, the observer in the inhomogeneous Universe sees
the same orientation of the lens as the corresponding observer in the
homogeneous Universe sees of the equivalent single-plane lens. Since
the image positions $\vc\theta^i$ are the same in both cases, this
implies that the positions of the images relative to the observed
orientation of the lens are the same in both cases, in contrast to the
conjecture made in Blandford \& Kundi\'c (1996). In fact, not only is
the observed orientation the same in both situations, but the
potentials do agree in both cases when expressed in terms of
$\vc\theta$! Note that this is not the case for the surface mass
densities.

Finally, it can be easily seen that if one considers a family of
models where the primary lens has elliptical
potential curves and in addition a shear matrix is included, then 
the cosmological contributions $C_\theta$ and $\R_{\rm d}$ yield an
equivalent single-plane lens which is included within this family.

\sec{6 Summary and discussion}
In this paper the effects of large-scale matter inhomogeneities on
gravitational lensing by galaxies were considered. The lens equation in
an inhomogeneous Universe was derived; the only approximation entering
(2.9) is that the length scale of the matter inhomogeneities is
much larger than the linear size of a light bundle which encloses all
light rays from the source to the observer (i.e., for a typical
multiply imaged QSO, this size is of order 20\ts kpc). In that case,
light propagating from the source to the lens and from the lens to the
observer can be described by the Jacobi differential equation. Note
that for the validity of (2.9) one does not need to assume that the
inhomogeneities are weak -- or in other words, that the 
matrices $\R_{\rm d}$, $\R_{\rm s}$ and $\R_{\rm ds}$ are close to the
unit matrix. The analogy of the `telescope matrix' which was defined
by Kovner (1987) has been shown to be symmetric in general. This
symmetry property is essential for showing that for arbitrary
propagation matrices $\R_{\rm d}$, $\R_{\rm s}$ and $\R_{\rm ds}$
caused by the large-scale structure, there exists an equivalent
single-plane gravitational lens such that all observables are the same
-- image positions, relative magnification matrices and
time-delays. However, in general this equivalent single-plane lens
contains a homogeneous matter sheet, not necessarily with positive
surface mass density. Therefore, the mass-sheet degeneracy as
discussed in Gorenstein, Falco \& Shapiro (1988) does apply. 

For a single lens plane, this mass-sheet degeneracy can be broken if
independent observational information about the lens can be
obtained. For example, if the lens is well described by an isothermal
mass distribution, the measurement of the velocity dispersion
determines the lens strength and thus fixes the mass-sheet
transformation. If this is done, the time delay then depends on the
matrix $\R_{\rm d}$, and thus matter inhomogeneities between us and
the lens affect the determination of the Hubble constant. This fact
was noted by Narayan (1991) who showed that the measurement of the
time delay determines the angular diameter distance to the lens, which
is affected by $\R_{\rm d}$, in agreement with B-K96, where the size
of this effect has been calculated using the non-linear evolution of
the power spectrum. Thus, lensing by the large-scale structure does
affect the determination of the Hubble constant, although the
magnitude of the effect (i.e., the rms deviation of $\R_{\rm d}$ from the
unit matrix) is of the order of a few percent; see Fig.\ts 2 of
B-K96. Therefore, in agreement with the conclusion of Surpi, Harari \&
Frieman (1996), cosmic shear does not seriously compromise the lensing
method for the determination of $H_0$.

Whereas the mass distribution of the equivalent single-plane lens
depends non-trivially on the lens mass distribution and the
propagation matrices $\R$, it has been shown that if the lens is
chosen to have elliptical isopotentials then the equivalent
single-plane lens consists of a main component with elliptical
isopotentials plus a shear matrix. Furthermore, if a family of lens
models is considered, consisting of an elliptical isopotential
component and a shear matrix, then the equivalent single-plane lens is
also within this family. For this case it was shown that the image
positions relative to the orientation of the mass distribution as seen
by an observer is the same in the inhomogeneous Universe and in the
single-plane lensing situation. Note that lens models consisting of an
elliptical lens (either elliptical isodensity contour or elliptical
isopotentials) plus external shear seems to be required for those lens
systems in which there is a large number of observational constraints
(e.g., Keeton, Kochanek \& Seljak 1996, Witt \& Mao 1997).

I thank S.\ts White, J.\ts Frieman, and in particular S.\ts Seitz for
valuable discussions, and M.\ts Bartelmann for comments on the
manuscript. \SFB
\vfill\eject

\sec{Appendix}
We shall show in this Appendix that the matrix $C_\xi$ multiplying
$\vc\xi$ in (2.11) is symmetric. To do this, we define the matrix
$$
A(\lambda):=\D(\lambda)\D^{-1}(\lambda_{\rm d})
\eqno (A1)
$$
and the matrix
$$
C(\lambda)=\hat\D^{-1}(\lambda)\D(\lambda)\D^{-1}(\lambda_{\rm d})
=\hat\D^{-1}(\lambda)A(\lambda)\quad ,
\eqno (A2)
$$
in both cases for $\lambda\ge\lambda_{\rm d}$. The proof proceeds in
the following four steps:

(1) The matrix $A(\lambda)$ satisfies the differential equation (2.3),
which implies that $A$ and $\hat D$ satisfy the same differential
equation. This can be trivially shown by inserting (A1) into (2.3).

(2) For any two solutions $X$ and $Y$ of the differential equation
(2.3), one finds
$$
{\d\over \d\lambda}\eckk{(X^{\rm T})'  Y-X^{\rm T} Y'} =0\quad ,
\eqno (A3)
$$
as can be easily verified by straight insertion into (2.3). This
relation implies that 
$$
S(\lambda):=\hat\D'(\lambda) \hat\D^{-1}(\lambda)
\eqno (A4)
$$
is symmetric. This can be shown as follows: Eq.\ts(A3) implies, by
setting $X=Y=\hat\D$, that $(\hat\D^{\rm T})'\hat\D-\hat\D^{\rm
T}\hat\D'$ is constant. Due to the initial condition at
$\lambda=\lambda_{\rm d}$, $\hat\D^{\rm T}(\lambda_{\rm
d})=0=\hat\D(\lambda_{\rm d})$, this constant is the zero matrix, 
so that $(\hat\D^{\rm
T})'\hat\D=\hat\D^{\rm T}\hat\D'$ for all $\lambda\ge\lambda_{\rm
d}$. In particular this implies
$$
(\hat\D^{\rm T})'=\hat\D^{\rm T}\hat\D' \hat\D^{-1}\quad .
\eqno (A5)
$$
To show the symmetry of $S$, we consider
$$
S-S^{\rm T}=\hat\D' \hat\D^{-1}
-\rund{\hat\D^{-1}}^{\rm T}(\hat\D^{\rm T})'
= \hat\D' \hat\D^{-1} -\rund{\hat\D^{-1}}^{\rm T}
\hat\D^{\rm T}\hat\D' \hat\D^{-1} =0\quad ,
\eqno (A6)
$$
where we have used (A5) in the second step, and in the final step the
fact was
employed that the inverse and transpose operations do commute. 
Hence, $S$ is symmetric.

(3) If we set $X=A$ and $Y=\hat\D$, we find that the constant in the
bracket of (A3) is
$$
(A^{\rm T})'\hat\D-A^{\rm T}\hat\D'=-(1+z_{\rm d})\I\quad ,
\eqno (A7)
$$
due to the intial conditions $A(\lambda_{\rm d})=\I$,
$\hat\D(\lambda_{\rm d})=0$, $\hat\D'(\lambda_{\rm d})=(1+z_{\rm
d})\I$. Hence,
$$
(1+z_{\rm d})\hat\D^{-1}=A^{\rm T} S-(A^{\rm T})'\quad .
\eqno (A8)
$$

(4) The final step is taken by noting that
$\lim_{\lambda\to\lambda_{\rm d}}\eck{(\lambda-\lambda_{\rm
d})C(\lambda)}$ is symmetric. We now show that $C'$ is symmetric for
all $\lambda \ge\lambda_{\rm d}$, which then proves that $C$ is
symmetric for all $\lambda -\lambda_{\rm d}$, and in particular this
is true for $C_\xi=C(\lambda_{\rm s})$. Using 
$$
\hat\D \hat\D^{-1}=\I \to
(\hat\D^{-1})'=-\hat\D^{-1} S\quad,
$$
as follows from differentiation,
we find in turn:
$$\eqalign{
C'&=\hat\D^{-1} A' + (\hat\D^{-1})' A
=\hat\D^{-1} A'- \hat\D^{-1} S A\cr
&={1\over (1+z_{\rm d})}\rund{A^{\rm T} S A' 
-(A^{\rm T})'A'-A^{\rm T}S S A
+ (A^{\rm T})'S A} \quad .\cr }
\eqno (A9)
$$
However, this final expression is manifestly symmetric, which
completes the proof. 

\def\ref#1{\vskip1pt\noindent\hangindent=40pt\hangafter=1 {#1}\par}
\sec{References}
\ref{Bar-Kana, R.\ 1996, ApJ 468, 17 (B-K96).}
\ref{Bartelmann, M. \& Schneider, P.\ 1991, A\&A 248, 349.}
\ref{Bernardeau, F., van Waerbeke, L. \& Mellier, Y.\ 1996,
astro-ph/9609122.} 
\ref{Blandford, R.D. \& Kochanek, C.S.\ 1987, ApJ 321, 658.}
\ref{Blandford, R.D. \& Kundi\'c, T.\ 1996, astro-ph9611229.}
\ref{Blandford, R.D., Saust, A.B., Brainerd, T.G. \& Villumsen, J.V.\
1991, MNRAS 251, 600.}
\ref{Cooke, J.H. \& Kantowski, R.\ 1975, ApJ 195, L11.} 
\ref{Gorenstein, M.V., Falco, E.E. \& Shapiro, I.I.\ 1988, ApJ 327,
693.} 
\ref{Gunn, J.E.\ 1967a, ApJ 147, 61.}
\ref{Gunn, J.E.\ 1967b, ApJ 150, 737.}
\ref{Jain, B. \& Seljak, U.\ 1996, astro-ph/9611077.}
\ref{Jaroszy\'nski, M., Park, C., Paczy\'nski, B. \& Gott, J.R.\ 1990, ApJ
365, 22.}
\ref{Jaroszy\'nski, M.\ 1991, MNRAS 249, 430.}
\ref{Jaroszy\'nski, M.\ 1992, MNRAS 255, 655.}
\ref{Kaiser, N.\ 1992, ApJ 388, 272.}
\ref{Kaiser, N.\ 1996, astro-ph/9610120.}
\ref{Kassiola, A. \& Kovner, I.\ 1993, ApJ 417, 450.}
\ref{Kayser, R. \& Refsdal, S.\ 1983, A\&A 128, 156.}
\ref{Keeton, C.R., Kochanek, C.S. \& Seljak, U.\ 1996, astro-ph/9610163.}
\ref{Kovner, I.\ 1987, ApJ 316, 52.}
\ref{Kovner, I.\ 1990, ApJ 351, 114.}
\ref{Miralda-Escud\'e, J.\ 1991, ApJ 380, 1.}
\ref{Narayan, R.\ 1991, ApJ 378, L5.}
\ref{Schneider, P.\ 1984, A\&A 140, 119.}
\ref{Schneider, P.\ 1985, A\&A 143, 413.}
\ref{Schneider, P., Ehlers, J. \& Falco, E.E.\ 1992, {\it
Gravitational Lenses}, Springer-Verlag (SEF).}
\ref{Seitz, S. \& Schneider, P.\ 1992, A\&A 265, 1.}
\ref{Seitz, S. \& Schneider, P.\ 1994, A\&A 287, 349.}
\ref{Seitz, S., Schneider, P. \& Ehlers, J.\ 1994,
Class. Quant. Grav. 11, 2345 (SSE).}
\ref{Seljak, U.\ 1996, ApJ 463, 1.}
\ref{Surpi, G.C., Harari, D.D. \& Frieman, J.A.\ 1996, ApJ 464, 54.}
\ref{Villumsen, J.\ 1996, MNRAS 281, 369.}
\ref{Wambsganss, J., Cen, R. \& Ostriker, J.P.\ 1996,
astro-ph/9610096.}
\ref{Witt, H.-J. \& Mao, S.\ 1997, preprint.}
\vfill\eject\end